\documentclass[aps,prd,twocolumn,showpacs,floatfix,preprintnumbers,amsfont,amsmath,amssymb,nofootinbib, superscriptaddress]{revtex4-1}

\usepackage{graphicx}
\usepackage{dcolumn}
\usepackage{bm}
\usepackage{epsfig}
\usepackage{amsmath}
\usepackage{xcolor}

\begin{document}

\title{Gravitational Waves from Newborn Accreting Millisecond Magnetars}

\author{Shu-Qing Zhong}
\affiliation{School of Astronomy and Space Science, Nanjing University, Nanjing 210093, China}
\affiliation{Key laboratory of Modern Astronomy and Astrophysics (Nanjing University), Ministry of Education, Nanjing 210093, China}

\author{Zi-Gao Dai}
\email{dzg@nju.edu.cn}
\affiliation{School of Astronomy and Space Science, Nanjing University, Nanjing 210093, China}
\affiliation{Key laboratory of Modern Astronomy and Astrophysics (Nanjing University), Ministry of Education, Nanjing 210093, China}

\author{Xiang-Dong Li}
\affiliation{School of Astronomy and Space Science, Nanjing University, Nanjing 210093, China}
\affiliation{Key laboratory of Modern Astronomy and Astrophysics (Nanjing University), Ministry of Education, Nanjing 210093, China}

\begin{abstract}
Two accretion columns have been argued to form over the surface of a newborn millisecond magnetar for an extremely high accretion rate $\gtrsim1.8\times10^{-2}M_\odot\ {\rm s^{-1}}$ that may occur in the core-collapse of a massive star. In this paper, we investigate the characteristics of these accretion columns and their gravitational wave (GW) radiation. For a typical millisecond magnetar (surface magnetic field strength $B\sim10^{15}~$G and initial spin period $P\sim1~$ms), we find (1) its accretion columns are cooled via neutrinos and can reach a height $\sim1$~km over the stellar surface; (2) its column-induced characteristic GW strain is comparable to the sensitivities of the next generation ground-based GW detectors within a horizon $\sim1$~Mpc; (3) the magnetar can survive only a few tens of seconds; (4) during the survival timescale, the height of the accretion columns increases rapidly to the peak and subsequently decreases slowly; (5) the column mass, characteristic GW strain, and maximum GW luminosity have simultaneous peaks in a similar rise-fall evolution. In addition, we find that the magnetar's spin evolution is dominated by the column accretion torque. A possible association with failed supernova is also discussed.

\end{abstract}

\maketitle

\section{Introduction}
\label{sec:introduction}

Since the gravitational wave (GW) event GW150914 opened a new window to observe the Universe \cite{abb16},
ten other GW events have been also detected by the advanced LIGO/Virgo detectors in the observing run 1 \& 2 (O1 \& O2) \cite{abb19a}. Among them, GW170817
is a watershed event from the merger of neutron star--neutron star (NS--NS),
having initiated electromagnetic wave-gravitational wave (EM-GW) multi-messenger astronomy \cite{abb17a,abb17b}.
Although its remnant, being a magnetar or a black hole (BH),
remains controversial \cite{abb17a,abb17c,abb17d,yu18}.

Theoretically, newborn millisecond magnetars are widely thought to be rapidly rotating NSs
with extremely strong magnetic fields \cite{dun92,thom93,dai98,dai04,dai12,lv14,lv15},
which are produced after the mergers of binary NSs or the core collapses of massive stars, accompanied by gamma-ray bursts (GRBs).
Speculatively, these magnetars are generally deemed to generate a strong X-ray emission (flare or plateau)
in their GRB afterglows \cite{dai06,tro07,row13,lv14},
and strong GW radiation effect on X-ray plateau
due to a large ellipticity and a high spin \cite{sha83,zhang01,cor09,fan13}.
This type of strong GW radiation should be detectable with the advanced LIGO/Virgo detectors and the
Einstein Telescope (ET) within a horizon of $\sim100$\,Mpc in the future, together with an X-ray plateau \cite{las16,lv18}.
In addition, other asymmetric deformations such as magnetic flare-induced oscillation \cite{kas11,zin12},
rotation-induced radial oscillation \cite{dai19}, and accretion instability-induced non-axisymmetric shape \cite{piro12} also generate strong GW radiation.

In an accreting magnetar scenario, a different structural asymmetry is caused by accreting materials
that are magnetically channeled towards the magnetar polar caps and become two columns,
the so-called accretion columns \cite{fra92}. For the NSs in X-ray binaries,
their accretion column-induced deformation generates too weak GW radiation to be detected by the
advanced LIGO/Virgo detectors and the next generation GW detectors \cite{kon16}. However,
the accretion columns of a millisecond magnetar, formed under some extreme conditions such as an extremely high accretion rate,
should make the columns significantly different from those of a normal NS in an X-ray binary. This resultant deformation could also lead to strong GW radiation.

In this paper, we first investigate the properties of the accretion columns of a newborn millisecond magnetar, the column-induced GW emission, and their time evolutions. We then calculate the magnetar's spin evolution affected by the column accretion, column-induced GW radiation, and magnetic dipole radiation, and speculate possible association between the column accretion of a magnetar and a failed supernova.
It should be noted that a magnetar can be formed from several channels
such as the core-collapse of a massive star \cite{whe00,buc08,buc09},
the merger of two NSs \cite{ros03,dai06,met08,gia13} or two white dwarfs (WDs) \cite{yoo07} and the accretion-induced collapse (AIC) of a WD \cite{tau13,sch15}. In this work, we just focus on the accretion columns formed over a millisecond magnetar for an extremely high accretion rate that may occur in the core-collapse of a massive star.

\section{Accretion Columns and GW Radiation}
\label{sec:column and GW}

\begin{figure}
\includegraphics[width=0.47\textwidth, angle=0]{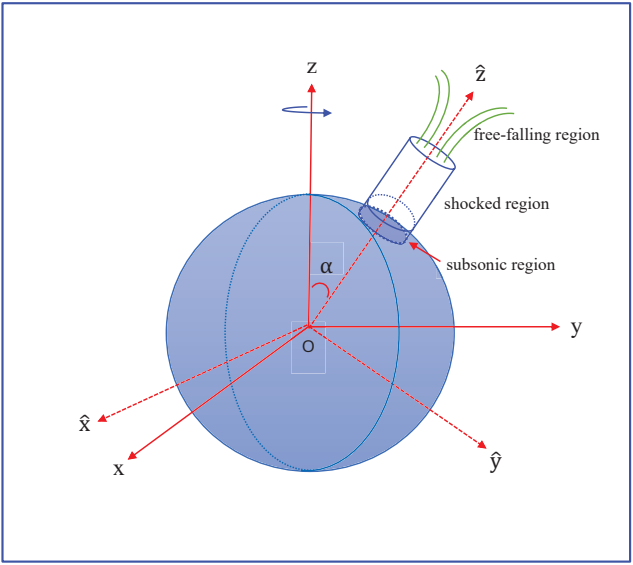}
\caption{Schematic picture of a magnetar accretion column along the $\hat{z}$ magnetic axis
with the misalignment angle $\alpha$ relative to the rotational axis $z$.
The material flow consists of a free-falling region, a high and dominantly-shocked region,
and a possibly thin subsonic region in which the flow settles onto the stellar surface. The gravitational potential energy of accreting material would be released through neutrino cooling.}
\label{fig:column}
\end{figure}

We consider a newborn millisecond magnetar that accretes the fall-back material immediately after its formation.
Its extremely strong magnetic field ($\sim10^{14}-10^{15}$ G) and millisecond period
\cite{thom93,lv14} probably make its
accretion and outflow very different from those of a normal NS.
In the case of $R<r_{\rm m}<r_{\rm c}$
(where $R$, $r_{\rm m}$, and $r_{\rm c}$ are the magnetar's radius,
Alfv\'{e}n radius, and co-rotation radius, respectively),
the fall-back material is magnetically channeled into the magnetar's
dipole caps and accumulated into two flowing accretion columns.
Each material flow consists of a free-falling region, a high and dominantly-shocked region,
and a possibly thin subsonic region in which the flow settles onto the stellar surface
\cite{bas76,lyu88}.
When the accretion, outflow, and settling are in hydrostatics equilibrium,
the accreting material would finally settle onto the star and become a part of the star,
and its large fraction of gravitational potential energy
would be liberated through neutrino cooling by considering electron-positron pair annihilation rather than photon cooling
due to a long diffusion timescale of photons $\sim10^4$ s
before it reaches the star's surface \cite{piro11}.
The magnetar has an angular velocity $\Omega$ along the $z$ rotational axis,
usually with the misalignment angle $\alpha$ relative to the accretion columns along $\hat{z}$ magnetic axis,
as shown in Figure \ref{fig:column}.

Based on \cite{piro11},
the magnetar's Alfv\'{e}n radius $r_{\rm m}$ is written as
\begin{eqnarray}
r_{\rm m} &=& \mu^{4/7}(GM)^{-1/7}\dot{M}^{-2/7}\nonumber\\
&=& 14\mu_{33}^{4/7}M_{1.4}^{-1/7}\dot{M}_{-2}^{-2/7}\ {\rm km}\nonumber\\
&=& 20B_{15}^{4/7}R_{12}^{12/7}M_{1.4}^{-1/7}\dot{M}_{-2}^{-2/7}\ {\rm km},
\label{eq:rm}
\end{eqnarray}
where $\mu\equiv BR^3$, $M$, and $\dot{M}$ are the magnetar's dipole magnetic moment, mass,
and accretion rate, respectively;
and the surface magnetic field strength $B_{15}=B/10^{15}~{\rm G}$, the magnetar radius $R_{12}=R/12\ {\rm km}$,
$M_{1.4}=M/1.4~M_\odot$, and $\dot{M}_{-2}=\dot{M}/10^{-2}~M_\odot\ {\rm s^{-1}}$.
Another critical radius is the co-rotation radius $r_c$,
\begin{eqnarray}
r_{\rm c} = (GM/\Omega^2)^{1/3} = 17M_{1.4}^{1/3}P_1^{2/3}\ {\rm km},
\label{eq:rc}
\end{eqnarray}
where $P_1=P/1~{\rm ms}=2\pi/\Omega/1~{\rm ms}$.
When an accretion column forms, its area at a radius $r$
can be scaled as
\begin{eqnarray}
    A(r)\approx \pi r^2\sin^2\theta \approx \pi r^3/r_{\rm m}.
	\label{eq:Ar}
\end{eqnarray}
After free-falling, the material will be decelerated and heated in the shocked region.
We consider the neutrino cooling primarily via electron-positron pair annihilation in this region, with a luminosity density \cite{pop99}
\begin{eqnarray}
	\dot{q}_{\rm pairs} = 5\times10^{33}T_{\rm sh,11}^9\ {\rm erg\ cm^{-3}\ s^{-1}},
	\label{eq:qparis}
\end{eqnarray}
where $T_{\rm sh, 11}=T_{\rm sh}/10^{11}\ {\rm K}$
and $T_{\rm sh}=[(9\sqrt{2GM})\dot{M}r_{\rm m}/(\pi ar^{7/2})]^{1/4}$ ($a$ is the radiation energy density constant)
obtained from equations
(A2), (A7), (A9) in \cite{piro11}, and equation (\ref{eq:Ar}) above.
According to the gravitational potential energy radiative luminosity $GM\dot{M}/R$ and
the energy conservation of hydrostatics equilibrium, we can obtain
\begin{eqnarray}
	GM\dot{M}/R &=& \int_R^{R+h_{\rm ac}}A(r)\dot{q}_{\rm pairs}(r)dr \nonumber\\
&\approx& A(R)\dot{q}_{\rm pairs}(R)h_{\rm ac},
    \label{eq:g=qp}
\end{eqnarray}
where we consider the height of neutrino
cooling over the magnetar surface as the height of the accretion column $h_{\rm ac}$ and assume $h_{\rm ac}\ll R$.
It can be estimated from equations (\ref{eq:rm}),
(\ref{eq:Ar}), (\ref{eq:qparis}), and (\ref{eq:g=qp}) (its step-by-step derivation can be seen in Appendix \ref{sec:derivation}),
\begin{eqnarray}
	h_{\rm ac}&\approx&\frac{GM\dot{M}R^{-1}}{A(R)\dot{q}_{\rm pairs}(R)} \nonumber \\
&=& 2\times10^{65} \pi^{5/4}\left(\frac{a}{9\sqrt{2}}\right)^{9/4}(GM)^{3/56}\dot{M}^{-25/28}\nonumber \\
&\times&B^{-5/7}R^{97/56}\ ({\rm erg\ cm^{-3}\ s^{-1}\ K^{-9}})^{-1}
	\nonumber
	\\
	&=& 8.8\times10^{4}M_{1.4}^{3/56}\dot{M}_{-2}^{-25/28}B_{15}^{-5/7}R_{12}^{97/56}~{\rm cm},
    \label{eq:h_ac}
\end{eqnarray}
This is a natural result that the column height is comparable to the shock height estimated
by equation (A15) in \cite{piro11}
because the neutrino cooling is exactly induced by the shock heating.

Using equations (A2) and (A7) in \cite{piro11} and equation (\ref{eq:Ar}) above,
we can get the density $\rho_{\rm sh}=7\dot{M}r_{\rm m}/[2\pi (2GM)^{1/2}r^{5/2}]$
and its corresponding mass for one accretion column (also assuming $h_{\rm ac}\ll R$,
its derivation also can be seen in Appendix \ref{sec:derivation})
\begin{eqnarray}
	M_{\rm ac}&=&\int_R^{R+h_{\rm ac}}A(r)\rho_{\rm sh}(r)dr
	\nonumber
	\\
    &\approx& \frac{7}{2\sqrt{2}}\dot{M}(GM)^{-1/2}R^{1/2}h_{\rm ac}
    \nonumber
	\\
    &=&  \frac{7}{\sqrt{2}}\times10^{65} \pi^{5/4}\left(\frac{a}{9\sqrt{2}}\right)^{9/4}(GM)^{-25/56}\dot{M}^{3/28}\nonumber \\
    &\times&B^{-5/7}R^{125/56}~({\rm erg\ cm^{-3}\ s^{-1}\ K^{-9}})^{-1}
    \nonumber
	\\
    &=& 1.7\times10^{-7}M_{1.4}^{-25/56}\dot{M}_{-2}^{3/28}B_{15}^{-5/7}R_{12}^{125/56}~{M_\odot},
	\label{eq:M_ac}
\end{eqnarray}

We suppose that there is the misalignment angle $\alpha$ between the rotational axis
and the accretion column along magnetic axis
so as to generate GW emission \cite{bon96}.
In the case of rigid rotation about a nonprincipal axis \cite{sha83},
the mass quadrupole moment can be represented as
$Q=I_{\hat{z}\hat{z}}-I_{\hat{x}\hat{x}}=(M+2M_{\rm ac})(R+h_{\rm ac})-MR^2$, so we find
\begin{eqnarray}
	Q &\approx& 2M_{\rm ac}R^2
    \nonumber
	\\
    &=& \frac{14}{\sqrt{2}}\times10^{65} \pi^{5/4}\left(\frac{a}{9\sqrt{2}}\right)^{9/4}(GM)^{-25/56}\dot{M}^{3/28}\nonumber\\
    &\times&B^{-5/7}R^{237/56}~({\rm erg\ cm^{-3}\ s^{-1}\ K^{-9}})^{-1}
    \nonumber
	\\
    &=& 1.0\times10^{39}M_{1.4}^{-25/56}\dot{M}_{-2}^{3/28}B_{15}^{-5/7}R_{12}^{237/56}~{\rm g~cm^2}.
\end{eqnarray}
Therefore, the maximum amplitude of GW radiation produced from this mass quadrupole moment $Q$ at a distance $D$ of the source to the observer is given by \cite{bon96,kon16}
\begin{eqnarray}
	h_0 &=& \frac{24\pi^2G}{c^4P^2D}Q
    \nonumber \\
    &=& \frac{48\pi^2G}{c^4}P^{-2}D^{-1}M_{\rm ac}R^2.
\end{eqnarray}

If the misalignment angle $\alpha$ is small, the GW emission at frequency $f=\Omega/2\pi=1/P$ is dominant, corresponding to a maximum amplitude $\frac{1}{2}h_0$. For $\alpha\sim \pi/2$, the GW emission is primarily at $f=2/P$, corresponding to a maximum amplitude $h_0$. Otherwise, for $\alpha=0$ or $\pi$, there is no GW emission \cite{sha83,bon96}. In the following, we take the misalignment angle $\alpha=\pi/2$ as an instance since it represents an upper limit of GW emission.
In this case, the characteristic GW strain can be approximated by \cite{cor09}
\begin{eqnarray}
    h_c &=& fh_0\sqrt{\frac{dt_{\rm sur}}{df}}\simeq h_0\sqrt{ft_{\rm sur}}
    \nonumber
	\\
    &=& 1.4\times10^{-24}M_{1.4}^{-25/56}\dot{M}_{-2}^{3/28}\nonumber \\
    &\times&B_{15}^{-5/7}R_{12}^{237/56}P_1^{-2}D_{\rm Mpc}^{-1}.
    \label{eq:h_c}
\end{eqnarray}
where $D_{\rm Mpc}=D/1~{\rm Mpc}$, and $t_{\rm sur}\sim 25$ s
is the magnetar's survival timescale (see \S\,\ref{sec:time evoluton}). Given a fiducial magnetar parameter set
($M=1.4~M_{\odot}$, $B=10^{15}~$G, $R=12~$km, and $P=1~$ms),
the characteristic GW strain is comparable to the sensitivity of the next generation detector such as ET within
1 Mpc horizon (see Fig. 3 of \cite{las16} and refer to \cite{hil11}).

We also estimate the maximum GW luminosity, derived from equation (16.6.14) of \cite{sha83},
\begin{eqnarray}
	L_{\rm gw} &=& \frac{2G\Omega^6}{5c^5}Q^2\times16
    \nonumber
	\\
    &=& (2\pi)^6\frac{128G}{5c^5}P^{-6}M_{\rm ac}^2R^4
    \nonumber
	\\
    &=& 1.1\times10^{42}M_{1.4}^{-25/28}\dot{M}_{-2}^{3/14}\nonumber \\
    &\times&B_{15}^{-10/7}R_{12}^{237/28}P_1^{-6}~{\rm erg~s^{-1}},
    \label{eq:L_gw}
\end{eqnarray}
which is smaller than the magnetar's spin-down luminosity due to a magnetic dipole radiation
usually accounting for the X-ray plateau in GRB afterglow by a factor of several orders of magnitude for a fiducial magnetar parameter set.

\section{Time Evolution of Accretion Columns and GW Radiation}
\label{sec:time evoluton}

\begin{figure}
\includegraphics[width=0.44\textwidth, angle=0]{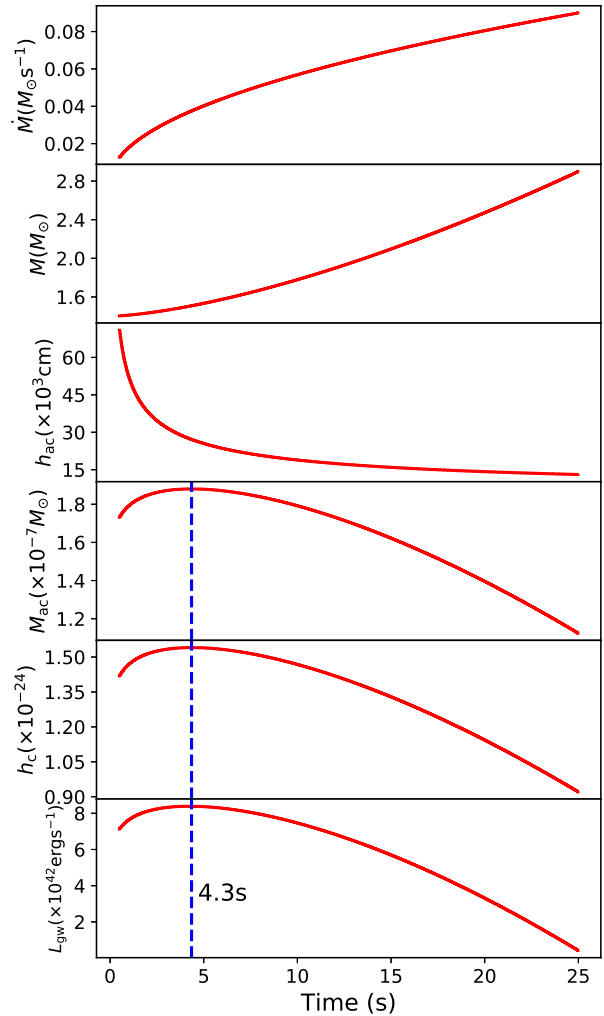}
\caption{The evolutions of accreting quantities and GW radiation.
As the accretion rate and magnetar mass increase,
the column height $h_{\rm ac}$ decreases rapidly to $t\sim4.3$ s and subsequently decreases slowly.
The column mass $M_{\rm ac}$, the characteristic GW strain $h_c$,
and the maximum GW luminosity $L_{\rm gw}$ have a similar rise-fall evolution.
They simultaneously reach peaks $\sim1.9\times10^{-7}~M_\odot$, $\sim1.5\times10^{-24}$ (within 1 Mpc horizon),
and $\sim8.4\times10^{42}~{\rm erg~s^{-1}}$ at $t\sim4.3$ s, respectively, under the fiducial magnetar parameter values.}
\label{fig:evolution}
\end{figure}

The accretion columns over the magnetar's surface occur in the case of $r_{\rm m}<r_{\rm c}$.
It implies a critical accretion rate
\begin{eqnarray}
\dot{M}>\dot{M}_{\rm cr} &=&1.8\times10^{-2}B_{15}^2R_{12}^6 \nonumber \\
&\times&M_{1.4}^{-5/3}P_1^{-7/3}\ M_{\odot}~{\rm s^{-1}}.
	\label{eq:mdotcrit}
\end{eqnarray}
This is a rather high accretion rate, which can be parameterized and assumed at early times
(see equation (14) of \cite{piro11} for mimicking the results of \cite{mac01,zhang08}),
\begin{eqnarray}
	\dot{M} = \eta 10^{-3}t_1^{1/2}M_\odot\ {\rm s^{-1}},
	\label{eq:mdotearly}
\end{eqnarray}
where $t_1=t/1~{\rm s}$ and $\eta$ is a parameter associated with supernova explosion energy
(a smaller $\eta$ corresponds to a larger explosion energy),
one can realize that $\eta$ should be very large
based on the critical accretion rate above.
Therefore, we adopt $\eta=18$, which exceeds
the normal range of $\eta\approx0.1-10$ mentioned in \cite{piro11}.
The reason that we adopt this value for $\eta$ is that the accretion rate should be larger than $\dot{M}_{\rm cr}=1.8\times10^{-2}\ M_{\odot}\ {\rm s^{-1}}$ after the start from $t=1\ {\rm s}$ from equations (\ref{eq:mdotcrit}) and (\ref{eq:mdotearly}) under the fiducial magnetar parameter values.
As we will discuss in \S\,\ref{subsec:failed supernova},
this extreme scenario might be associated with an extreme case---failed supernova \cite{koc08}.
For a magnetar with initial mass $M_0$, its time-dependent mass can be estimated from equation (\ref{eq:mdotearly}),
\begin{eqnarray}
M(t)&=&M_0+\int_0^{t}\dot{M}dt=M_0+1.2\times10^{-2}t_1^{3/2}~M_\odot  \nonumber\\
&=&M_0\left(1+\frac{1.2\times10^{-2}t_1^{3/2}~M_\odot}{M_0}\right)  \nonumber\\
&=&M_0\left(1+\frac{1.2\times10^{-2}t_1^{3/2}~M_\odot}{1.4~M_\odot}\right) \nonumber\\
&=&M_0(1+8.6\times10^{-3}t_1^{3/2})
\label{eq:mass}
\end{eqnarray}
In this case, the newborn millisecond magnetar survival timescale should be extremely short---only $t_{\rm sur}\sim25$ s
if $M_0=1.4~M_{\odot}$ and if the magnetar mass up to $2.9~M_{\odot}$ would collapse to a BH,
as shown in Figure \ref{fig:evolution}.

Combining equations (\ref{eq:mdotearly}) and (\ref{eq:mass})
and assuming that the term (i.e., $8.6\times10^{-3}t_1^{3/2}$) is smaller than unity in the final equality
of equation (\ref{eq:mass}) in the survival timescale of magnetar (i.e., $t_1<25$), we can estimate
$M^{\alpha}\dot{M}^{\beta} \sim M_0^{\alpha}(1+8.6\times10^{-3}{\alpha}t_1^{3/2})\times(1.8\times10^{-2}t_1^{1/2}M_{\odot}{\rm s^{-1}})^{\beta}=(1.4M_{\odot})^{\alpha}(1+8.6\times10^{-3}{\alpha}t_1^{3/2})\times(1.8\times10^{-2}M_{\odot}{\rm s^{-1}})^{\beta}(t_1^{\beta/2})$ (where $\alpha$ and $\beta$ are assumed to be power-law indexes of $M$ and $\dot{M}$, respectively).
In this way, we obtain the time evolutions of the following quantities and their corresponding results from equations (\ref{eq:h_ac}), (\ref{eq:M_ac}), (\ref{eq:h_c}), (\ref{eq:L_gw}), and the approximation of $M^{\alpha}\dot{M}^{\beta}$, as illustrated in Figure \ref{fig:evolution}.
\begin{eqnarray}
	h_{\rm ac} &\approx& 5.2\times10^{4}(1+4.6\times10^{-4}t_1^{3/2})t_1^{-25/56} \nonumber\\
&\times&B_{15}^{-5/7}R_{12}^{97/56}~{\rm cm},
\end{eqnarray}
\begin{eqnarray}
    M_{\rm ac} &\approx& 1.8\times10^{-7}(1-3.8\times10^{-3}t_1^{3/2})t_1^{3/56}  \nonumber\\
    &\times&B_{15}^{-5/7}R_{12}^{125/56}~{M_\odot},
\end{eqnarray}
\begin{eqnarray}
	h_c &\approx& 1.4\times10^{-24}(1-3.8\times10^{-3}t_1^{3/2})t_1^{3/56}  \nonumber\\
&\times&B_{15}^{-5/7}R_{12}^{237/56}P_1^{-2}D_{\rm Mpc}^{-1},
\end{eqnarray}
\begin{eqnarray}
	L_{\rm gw} &\approx& 7.7\times10^{42}(1-7.7\times10^{-3}t_1^{3/2})t_1^{3/28}  \nonumber\\
&\times&B_{15}^{-10/7}R_{12}^{237/28}P_1^{-6}~{\rm erg~s^{-1}}.
\end{eqnarray}
As the accretion rate and magnetar mass increase,
the column height $h_{\rm ac}$ decreases rapidly to $t\sim4.3$ s and subsequently decreases slowly.
Moreover, the column mass $M_{\rm ac}$, the characteristic GW strain $h_c$,
and the maximum GW luminosity $L_{\rm gw}$ have a similar rise-fall evolution.
They simultaneously reach peaks $\sim1.9\times10^{-7}~M_\odot$, $\sim1.5\times10^{-24}$ (within 1 Mpc horizon),
and $\sim8.4\times10^{42}~{\rm erg~s^{-1}}$ at $t\sim4.3$ s, respectively,
for the fiducial magnetar parameter values such as $M_0=1.4~M_{\odot}$, $B=10^{15}~$G, $R=12~$km, and $P=1~$ms.
In addition, it should be noted that these results are obtained by assuming that the magnetar's spin (or period $P$) is not time-dependent.

\section{Discussion}
\label{sec:discussion}

\subsection{Magnetar's Spin Evolution}
\label{subsec:spin evolution}

The magnetar's spin evolution should be affected by
the column accretion and its induced GW radiation, magnetic dipole radiation, and neutrino-driven wind radiation, when the misalignment angle $\alpha=\pi/2$.
For simplicity, we leave out the neutrino-driven wind radiation attribution to the magnetar spin as done in \cite{piro11}.
If the magnetar is regarded as a uniformly rotating body,
its angular momentum is removed (i.e., spin-down) via GW radiation at a rate
\begin{eqnarray}
	N_{\rm gw} &=& L_{\rm gw}/\Omega,
	\label{eq:ngw}
\end{eqnarray}
and via magnetic dipole radiation
\begin{eqnarray}
N_{\rm dip} &=& \frac{\mu^2\Omega^3}{6c^3}.
\label{eq:ndip}
\end{eqnarray}
In contrast, in the accretion column scenario ($R<r_{\rm m}<r_{\rm c}$),
the magnetar spins up due to an accretion torque
\begin{eqnarray}
    N_{\rm acc}&=& n(\omega)(GMr_{\rm m})^{1/2}\dot{M},
    \label{eq:nacc}
\end{eqnarray}
where $n(\omega)=(1-\omega)$ is the dimensionless torque
and $\omega = \Omega/(GM/r_{\rm m}^3)^{1/2} = (r_{\rm m}/r_{\rm c})^{3/2}$ is the fastness parameter \cite{piro11}.
When $r_{\rm m}<r_{\rm c}$ (i.e., $\dot{M}>1.8\times10^{-2}~M_{\odot}~{\rm s^{-1}}$), we get $\omega<1$ and $N_{\rm acc}>0$.
Therefore, the spin evolution is expressed as
\begin{eqnarray}
	I\dot{\Omega} = -N_{\rm gw}-N_{\rm dip}+N_{\rm acc},
	\label{eq:spin_differential}
\end{eqnarray}
where $I=0.4MR^2$ is the moment of inertia for an incompressible NS \cite{lat01}. We solve equation (\ref{eq:spin_differential}) by combining
equations (\ref{eq:rm}), (\ref{eq:rc}), (\ref{eq:L_gw}), (\ref{eq:mdotearly}), (\ref{eq:mass}), (\ref{eq:ngw}), (\ref{eq:ndip}), (\ref{eq:nacc}), and $\dot{\Omega}=-2\pi \dot{P}/P^2$,
with typical magnetar parameters of the initial mass $M_0=1.4~M_{\odot}$, the
initial period $P_0=1~{\rm ms}$, the magnetic field $B=10^{15}~{\rm G}$, and the stellar radius $R=12~{\rm km}$.

\begin{figure}
\includegraphics[angle=0,width=0.47\textwidth]{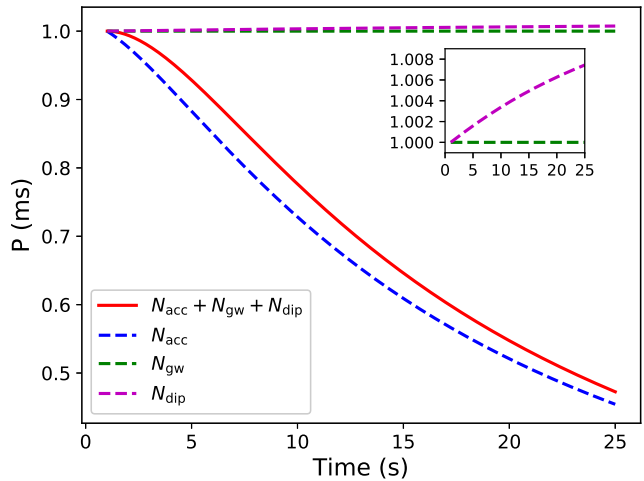}
\caption{The magnetar's spin evolution influenced by the spin-up accretion torque $N_{\rm acc}$,
and the spin-down GW radiation torque $N_{\rm gw}$ and dipole radiation torque $N_{\rm dip}$.
The red solid line represents the period evolution affected by all of these three torques.
The blue, green, and magenta dotted lines represent the period evolution by only
considering the accretion torque, GW radiation torque, and dipole radiation torque, respectively.
The spin evolution caused by the GW radiation torque and the dipole radiation torque is also displayed in the inset.}
\label{fig:spin}
\end{figure}

The results are plotted in Figure \ref{fig:spin}, in which the red solid line represents the period evolution
influenced by the spin-up accretion torque $N_{\rm acc}$,
and the spin-down GW radiation torque $N_{\rm gw}$ and dipole radiation torque $N_{\rm dip}$;
the blue, green, and magenta dotted lines represent the period evolution
by only considering the accretion torque, GW radiation torque, and dipole radiation torque, respectively.
As we can see, the accretion torque is dominant for the spin-up evolution of the magnetar
attributed to the accretion at a very high rate, in comparison to the GW radiation and dipole radiation
at a relatively low luminosity.
From the inset, it is shown that the dipole radiation luminosity is several orders of magnitude larger than the GW radiation luminosity.
Furthermore, before the magnetar collapses to a BH,
its period shortens from initial 1.00 ms to 0.40 ms. However, it is unclear whether and when the magnetar reaches its break-up limit. For a normal NS with mass $1.4~M_{\odot}$ and radius $12~{\rm km}$, its Keplerian break-up limit is $\sim$0.61 ms from equation (3) in \cite{lat04} for a rigid Newtonian sphere scenario. When considering deformation and General Relativity effects, the break-up limit of an NS is about $\frac{1}{1224{\rm Hz}}\sim$0.82 ms with the maximum-mass, nonrotating configuration \cite{lat04}, relying on its equation of state. For simplicity, we here don't consider the break-up limit of a millisecond magnetar.

\subsection{Possible Associated Failed Supernova}
\label{subsec:failed supernova}
As pointed out in \S\,\ref{sec:time evoluton}, the column accretion of a magnetar requires
a very high accretion rate corresponding to a very large $\eta$ at early times,
which suggests an extremely weak or even no supernova explosion---a failed supernova after the core-collapse of a massive star.
The failed supernova can well explain not only
the absence of high mass red supergiants ($16.5~M_\odot\leq M \leq25~M_\odot$)
as the progenitors of Type IIP SNe \cite{sma09a,sma09b},
but also the compact remnant mass gap between the observed NSs and
the observed BHs \cite{koc14} via both X-ray binaries \cite{oze10,far11}
and the recent GW events of BH--BH and NS--NS mergers \cite{abb19a,abb19b}.
Observationally, the failed supernova search has been made for many years \cite{koc08,ger15}
and has been confirmed from Large Binocular Telescope observations until recently \cite{ada17a,ada17b}.
In addition to a change from appearance to disappearance in a progenitor's electromagnetic radiation for a failed supernova,
we here surmise that the GW radiation induced by the magnetar's column accretion
is likely another association with the failed supernova,
especially because it would form initially a magnetar even that is temporary rather than a prompt BH
in the stellar core after the collapse of a massive star \cite{bur88,oo11}.
Additionally, this GW radiation may be another type of GW signature of core-collapse supernovae \cite{ott09}.
Nevertheless, we also see that this GW event lasts several tens of seconds
and its characteristic GW radiation strain is comparable to the sensitivities of
the next-generation GW detectors within 1~Mpc horizon.
Accordingly, it is unlikely to search for a failed supernova
by combining this short timescale GW radiation with the long timescale change
from appearance to disappearance in electromagnetic radiation.

\section{Conclusions}
\label{sec:conclusion}

We have investigated the properties of accretion columns of a newborn millisecond magnetar and the column accretion-induced GW radiation, and their evolutions. We have also discussed the magnetar's spin evolution by considering the effects of material
column accretion, GW radiation, and magnetic dipole radiation, and speculated the potential relation between this induced GW radiation and a failed supernova. Some interesting results are found as follows.
\begin{itemize}
\item For the fiducial magnetar parameter values
($M=1.4~M_{\odot}$, $B=10^{15}~$G, $R=12~$km, and $P=1~$ms), the magnetar's accretion columns are cooled via neutrinos,
its column height is relative to the shock height $\sim1~{\rm km}$;
the column mass is typically $\sim10^{-7}~M_\odot$,
the characteristic GW strain $\sim10^{24}$
is comparable to the sensitivities of the next generation GW detectors within 1~Mpc horizon;
and the maximum GW luminosity is $\sim10^{42}~{\rm erg~s^{-1}}$,
being smaller than the typical magnetar's magnetic dipole emission usually explaining
the observed X-ray plateaus in GRB afterglows by several orders of magnitude.

\item As the accretion rate increases at early times,
the magnetar mass also increases. The magnetar's survival
timescale is only about 25 s before collapsing to a BH for the fiducial magnetar parameter values mentioned above.
The column height rapidly declines before $t=4.3~{\rm s}$ and subsequently declines slowly.
The column mass, the characteristic GW strain, and the maximum GW luminosity have a similar rise-fall evolution
and simultaneous peaks at $t=4.3~{\rm s}$.

\item Thanks to the effects of column accretion, GW radiation, and dipole radiation,
the magnetar's spin should evolve with time. During the magnetar's survival timescale,
the column accretion torque is dominant for the magnetar's spin and makes it spun-up.

\item The column accretion scenario needs a very high accretion rate
corresponding to an extremely weak or even no supernova explosion---a failed supernova.
Thus the induced GW radiation is likely associated with a failed supernova
but unlikely to be used for probing for a failed supernova
by combining with the changing electromagnetic observation from appearance to disappearance in the future.
\end{itemize}

Finally, we simply estimate the event rate of the GW radiation induced by the column accretion of newborn millisecond magnetars.
The event rate of overall millisecond magnetar formation was estimated to be just a few 10-100\,Gpc$^{-3}$yr$^{-1}$ \cite{nic17}.
This rate is estimated, based on the event rate of short gamma-ray bursts (SGRBs). However, the millisecond magnetars we discussed here are probably associated with long gamma-ray bursts (LGRBs) generated from the core-collapse of massive stars. Moreover, the event rate of LGRBs is just about one third of that of SGRBs \cite{nic17}, so the overall millisecond magnetars associated with LGRBs might be only one third of those associated with SGRBs.
If so, for a volume of $\sim1$~Mpc$^3$ with the current and upcoming GW detectors,
the event rate of this GW radiation would be extremely low,
even if assuming all millisecond magnetars give rise to this kind of accretion columns.

\section*{Acknowledgments}
We really appreciate the referee for the careful review and very helpful comments and suggestions that have helped us improve the presentation of the paper.
This work is supported by the National Key Research and Development Program of China (grant No. 2017YFA0402600) and the National Natural Science Foundation of China (grant No. 11573014 and 11833003).

\appendix

\section{Equation Derivations}
\label{sec:derivation}
\subsection{Equation (\ref{eq:h_ac})}
From equations (A2), (A7), (A9) in \cite{piro11}, and equations (\ref{eq:rm}) and (\ref{eq:Ar}) in current paper, the temperature in the shocked region is
\begin{eqnarray}
T_{\rm sh}&=&\left(\frac{3p_{\rm sh}}{a}\right)^{1/4}=\left(\frac{18\rho_0v_{\rm in}^2}{a}\right)^{1/4} \nonumber \\
&=&\left[\frac{18(\dot{M}/2A)v_{\rm in}}{a}\right]^{1/4}=\left(\frac{9\dot{M}v_{\rm in}}{Aa}\right)^{1/4} \nonumber \\
&=&\left[\frac{9\dot{M}(2GM)^{1/2}r^{-1/2}r_m}{\pi ar^3}\right]^{1/4}
\nonumber \\
&=&\left[\frac{9\dot{M}(2GM)^{1/2}r_m r^{-7/2}}{\pi a}\right]^{1/4}  \nonumber \\
&=&\left(\frac{9\sqrt{2}}{\pi a}\right)^{1/4}(GM)^{1/8}\dot{M}^{1/4}r_m^{1/4}r^{-7/8}
\nonumber \\
&=&\left(\frac{9\sqrt{2}}{\pi a}\right)^{1/4}(GM)^{1/8}\dot{M}^{1/4} \nonumber \\
&\times&[\mu^{1/7}(GM)^{-1/28}\dot{M}^{-1/14}]r^{-7/8}  \nonumber \\
&=&\left(\frac{9\sqrt{2}}{\pi a}\right)^{1/4}(GM)^{5/56}\dot{M}^{5/28}\mu^{1/7}r^{-7/8}.
\end{eqnarray}
From equation (\ref{eq:qparis}),  the luminosity density via electron-positron pair annihilation is
\begin{eqnarray}
\dot{q}_{\rm pairs} &=& 5\times10^{33}T_{\rm sh,11}^9\ {\rm erg\ cm^{-3}\ s^{-1}}   \nonumber \\
&=&5\times10^{33}\times10^{-99}T_{\rm sh}^9\ {\rm erg\ cm^{-3}\ s^{-1}\ K^{-9}}  \nonumber \\
&=&5\times10^{-66}\left(\frac{9\sqrt{2}}{\pi a}\right)^{9/4}(GM)^{45/56}\dot{M}^{45/28} \nonumber \\
&\times&\mu^{9/7}r^{-63/8}\ {\rm erg\ cm^{-3}\ s^{-1}\ K^{-9}}.
\end{eqnarray}
Therefore,
\begin{eqnarray}
A(r)\dot{q}_{\rm pairs}(r)&=&5\times10^{-66}\left(\frac{9\sqrt{2}}{\pi a}\right)^{9/4}(GM)^{45/56}\dot{M}^{45/28}\nonumber \\
&\times&\mu^{9/7}r^{-63/8}(\pi r^3 r_m^{-1})~{\rm erg\ cm^{-3}\ s^{-1}\ K^{-9}}\nonumber \\
&=&5\times10^{-66}\pi\left(\frac{9\sqrt{2}}{\pi a}\right)^{9/4}(GM)^{45/56}\dot{M}^{45/28} \nonumber \\
&\times&\mu^{9/7}r^{-39/8}[\mu^{-4/7}(GM)^{1/7}\dot{M}^{2/7}] \nonumber \\
&\times&{\rm erg\ cm^{-3}\ s^{-1}\ K^{-9}} \nonumber \\
&=&5\times10^{-66}\pi\left(\frac{9\sqrt{2}}{\pi a}\right)^{9/4}(GM)^{53/56}\dot{M}^{53/28} \nonumber \\
&\times&\mu^{5/7}r^{-39/8}~{\rm erg\ cm^{-3}\ s^{-1}\ K^{-9}}.
\end{eqnarray}
From equation (\ref{eq:g=qp}), the height of the accretion column $h_{\rm ac}$ is
\begin{eqnarray}
h_{\rm ac}&\approx&\frac{GM\dot{M}R^{-1}}{A(R)\dot{q}_{\rm pairs}(R)} \nonumber \\
&=&GM\dot{M}R^{-1}\frac{1}{5\pi}\times10^{66}\left(\frac{9\sqrt{2}}{\pi a}\right)^{-9/4}(GM)^{-53/56} \nonumber \\
&\times&\dot{M}^{-53/28}\mu^{-5/7}R^{39/8}~({\rm erg\ cm^{-3}\ s^{-1}\ K^{-9}})^{-1} \nonumber \\
&=&\frac{1}{5\pi}\times10^{66}\left(\frac{9\sqrt{2}}{\pi a}\right)^{-9/4}(GM)^{3/56}\dot{M}^{-25/28}B^{-5/7} \nonumber \\
&\times&R^{97/56}~({\rm erg\ cm^{-3}\ s^{-1}\ K^{-9}})^{-1}\  ({\rm from}\ \mu=BR^3)
\nonumber \\
&=&2\times10^{65} \pi^{5/4}\left(\frac{a}{9\sqrt{2}}\right)^{9/4}(GM)^{3/56}\dot{M}^{-25/28} \nonumber \\
&\times&B^{-5/7}R^{97/56}~({\rm erg\ cm^{-3}\ s^{-1}\ K^{-9}})^{-1}.
\end{eqnarray}
The unit of the luminosity $GM\dot{M}/R$ is $\rm erg~s^{-1}$ in cgs, $A(r)$ corresponds to $\rm cm^2$,
and $\dot{q}_{\rm pairs}(r)$ corresponds to $\rm erg~cm^{-3}~s^{-1}$. Thus the unit of $h_{\rm ac}$ should be $\rm cm$.

\subsection{Equation (\ref{eq:M_ac})}
Using equations (A2) and (A7) in \cite{piro11} and equation (\ref{eq:Ar}) in this paper, the postshock density is
\begin{eqnarray}
\rho_{\rm sh}=7\rho_0=\frac{7\dot{M}}{2Av_{\rm in}}.
\end{eqnarray}
Hence,
\begin{eqnarray}
A\rho_{\rm sh}=\frac{7\dot{M}}{2v_{\rm in}}=\frac{7\dot{M}}{2\sqrt{2}}(GM)^{-1/2}r^{1/2}.
\end{eqnarray}
In this case, the mass of one accretion column is
\begin{eqnarray}
M_{\rm ac}&=&\int_R^{R+h_{\rm ac}}A(r)\rho_{\rm sh}(r)dr  \nonumber \\
&=&\frac{7\dot{M}}{2\sqrt{2}}(GM)^{-1/2}\times\frac{2}{3}[(R+h_{\rm ac})^{3/2}-R^{3/2}]   \nonumber \\
&\approx&\frac{7\dot{M}}{2\sqrt{2}}(GM)^{-1/2}R^{1/2}h_{\rm ac}  \nonumber \\
&=&\frac{7\dot{M}}{2\sqrt{2}}(GM)^{-1/2}R^{1/2} \nonumber \\
&\times& 2\times10^{65} \pi^{5/4}\left(\frac{a}{9\sqrt{2}}\right)^{9/4}(GM)^{3/56}\dot{M}^{-25/28} \nonumber \\
&\times&B^{-5/7}R^{97/56}~({\rm erg\ cm^{-3}\ s^{-1}\ K^{-9}})^{-1}   \nonumber \\
&=&  \frac{7}{\sqrt{2}}\times10^{65} \pi^{5/4}\left(\frac{a}{9\sqrt{2}}\right)^{9/4}(GM)^{-25/56}\dot{M}^{3/28}  \nonumber \\
&\times&B^{-5/7}R^{125/56}~({\rm erg\ cm^{-3}\ s^{-1}\ K^{-9}})^{-1}.
\end{eqnarray}

\end{document}